\documentclass{vgtc}                          

\ifpdf
\pdfoutput=1\relax                   
\usepackage{graphicx}                
\DeclareGraphicsExtensions{.pdf,.png,.jpg,.jpeg} 
\else
\usepackage{graphicx}                
\DeclareGraphicsExtensions{.eps}     
\fi%

\graphicspath{{figures/}{pictures/}{images/}{./}} 

\usepackage{microtype}                 
\PassOptionsToPackage{warn}{textcomp}  
\usepackage{textcomp}                  
\usepackage{mathptmx}                  
\usepackage{times}                     
\usepackage{cite}                      
\usepackage{tabu}                      
\usepackage{booktabs}                  

\usepackage{soul} 
\usepackage{balance} 
\usepackage{todonotes}

\onlineid{0}

\vgtccategory{Research}

\vgtcinsertpkg

\title{ReconViguRation: Reconfiguring Physical Keyboards in Virtual Reality}

\author{Daniel Schneider$^{1}$ \thanks{e-mail: daniel.schneider@hs-coburg.de. The first three authors contributed equally to this work.} %
	\and Alexander Otte$^{1}$\thanks{e-mail: alexander.otte@hs-coburg.de} %
	\and Travis Gesslein$^{1}$\thanks{e-mail: travis.gesslein@hs-coburg.de}
	\and Philipp Gagel$^{1}$\thanks{e-mail: philipp.gagel@stud.hs-coburg.de}
	\and Bastian Kuth$^{1}$\thanks{e-mail: bastian.kuth@stud.hs-coburg.de}
	\and Mohamad Sham Damlakhi$^{1}$\thanks{e-mail: Mohamad-Shahm.Damlakhi@stud.hs-coburg.de}
	\and Oliver Dietz$^{1}$\thanks{e-mail: oliver.dietz@stud.hs-coburg.de}
	\and Eyal Ofek$^{2}$\thanks{e-mail: eyalofek@microsoft.com}
	\and Michel Pahud$^{2}$\thanks{e-mail: mpahud@microsoft.com}
	\setcounter{footnote}{0}
	\and Per Ola Kristensson$^{3}$\thanks{e-mail: pok21@cam.ac.uk}
	\and J{\"o}rg M{\"u}ller$^{4}$\thanks{e-mail: joerg.mueller@uni-bayreuth.de}
	\and Jens Grubert$^{1}$\thanks{e-mail: jens.grubert@hs-coburg.de}
}
\affiliation{\scriptsize $^{1}$Coburg University of Applied Sciences and Arts  $^{2}$Microsoft Research  \\ $^{3}$University of Cambridge  $^{4}$University of Bayreuth}

\teaser{
	\centering
	\includegraphics[width=0.9\columnwidth]{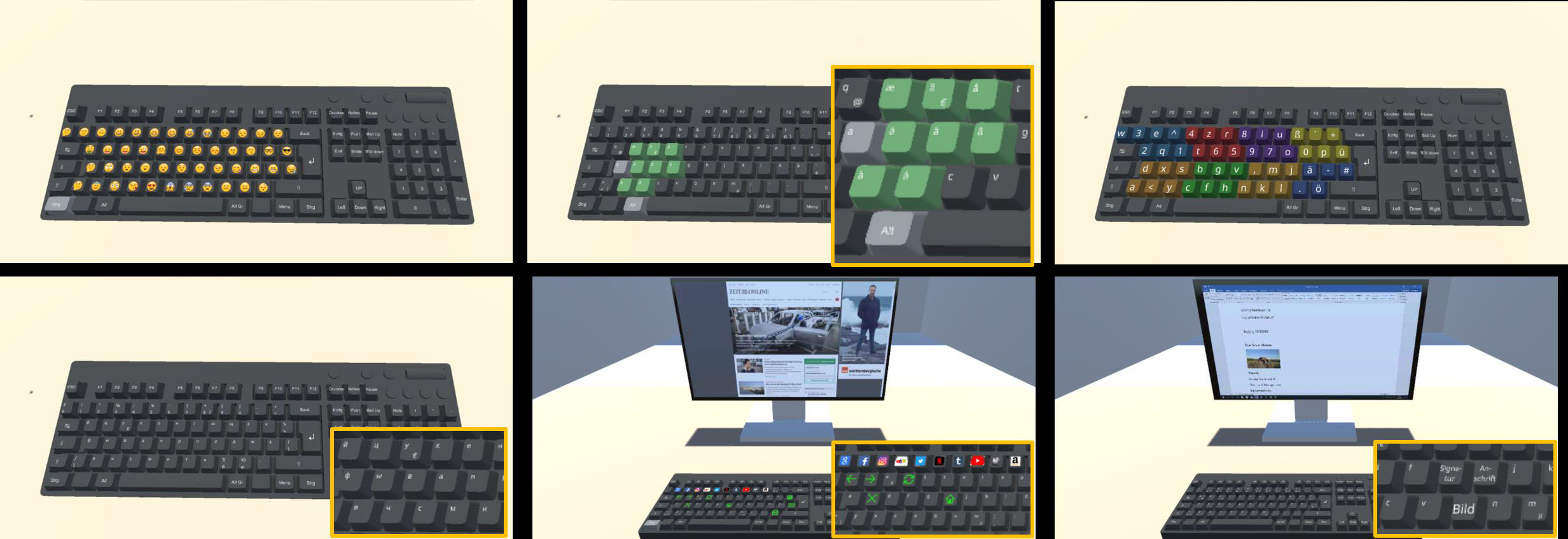}
	\caption{Example  for reconfiguring the input and output space of individual keys of a physical keyboard in virtual reality. Top row from left to right: emoji entry, special characters, secure password entry using randomized keys.
		Bottom row from left to right: foreign languages, browser shortcuts, text processing macros.}
	\label{fig:appskeys}
}

\abstract{
	Physical keyboards are common peripherals for personal computers and are efficient standard text entry devices. Recent research has investigated how physical keyboards can be used in immersive head-mounted display-based Virtual Reality (VR). So far, the physical layout of keyboards has typically been transplanted into VR for replicating typing experiences in a standard desktop environment.
	In this paper, we explore how to fully leverage the immersiveness of VR to change the input and output characteristics of physical keyboard interaction within a VR environment. This allows individual physical keys to be reconfigured to the same or different actions and visual output to be distributed in various ways across the VR representation of the keyboard.
	We explore a set of input and output mappings for reconfiguring the virtual presentation of physical keyboards and probe the resulting design space by specifically designing, implementing and evaluating nine VR-relevant applications: emojis, languages and special characters, application shortcuts, virtual text processing macros, a window manager, a photo browser, a whack-a-mole game, secure password entry and a virtual touch bar. We investigate the feasibility of the applications in a user study with 20 participants and find that, among other things, they are usable in VR. We discuss the limitations and possibilities of remapping the input and output characteristics of physical keyboards in VR based on empirical findings and analysis and suggest future research directions in this area.

} 

\CCScatlist{ 
	\CCScat{H.5.2:}{ User Interfaces - Input devices and strategies.}{}{}
}

\begin{document}

	
	\firstsection{Introduction}
	
	\maketitle


	Physical keyboards are common input peripherals, used for tasks ranging from text entry and editing to controlling PC games. Although there is a prevalence of touch keyboards in the mobile world, physical keyboards are still the best tool for text entry, rendering satisfying haptic feedback to the user and enabling fast interaction, even without a visual view of the keyboard \cite{Dhakal:2018:OTM:3173574.3174220}.
	
	Tracking users' hands and fingers allows the system to visualize the location of the fingers or hands in relation to the keyboard (e.g.~\cite{grubert2018text, grubert2018hand}). With the recent advances of hand and finger tracking technology (such as HoloLens 2 and Leap Motion), the industry is getting to a point where the use of physical keyboards in VR everywhere, such as in tiny touch-down spaces, is realizable without external tracking. This opens up new opportunities for pervasive VR applications, such as the VR office \cite{grubert2018office}, and motivates research in user interaction with physical keyboards in virtual reality.
	
	Recent research has demonstrated the superiority of physical keyboard for text entry in virtual applications, where the user is practically blind to the real world (e.g., \cite{mcgill2015dose, grubert2018text, knierim2018physical}). Most of these works represented the keyboard by a virtual model of similar geometry, carefully fit to lie at the exact location as the real keyboard.  VR is not compelled to follow the physical rules of our real world. Recent research has indicated that the location of the virtual and physical keyboard can differ without affecting text entry performance \cite{grubert2018text}. However, apart from changing the location of the virtual keyboard, further modifications of its visual representation or functionality has not been thoroughly investigated within VR.
	
\begin{figure}[tb]
		\centering
		\includegraphics[width=\columnwidth]{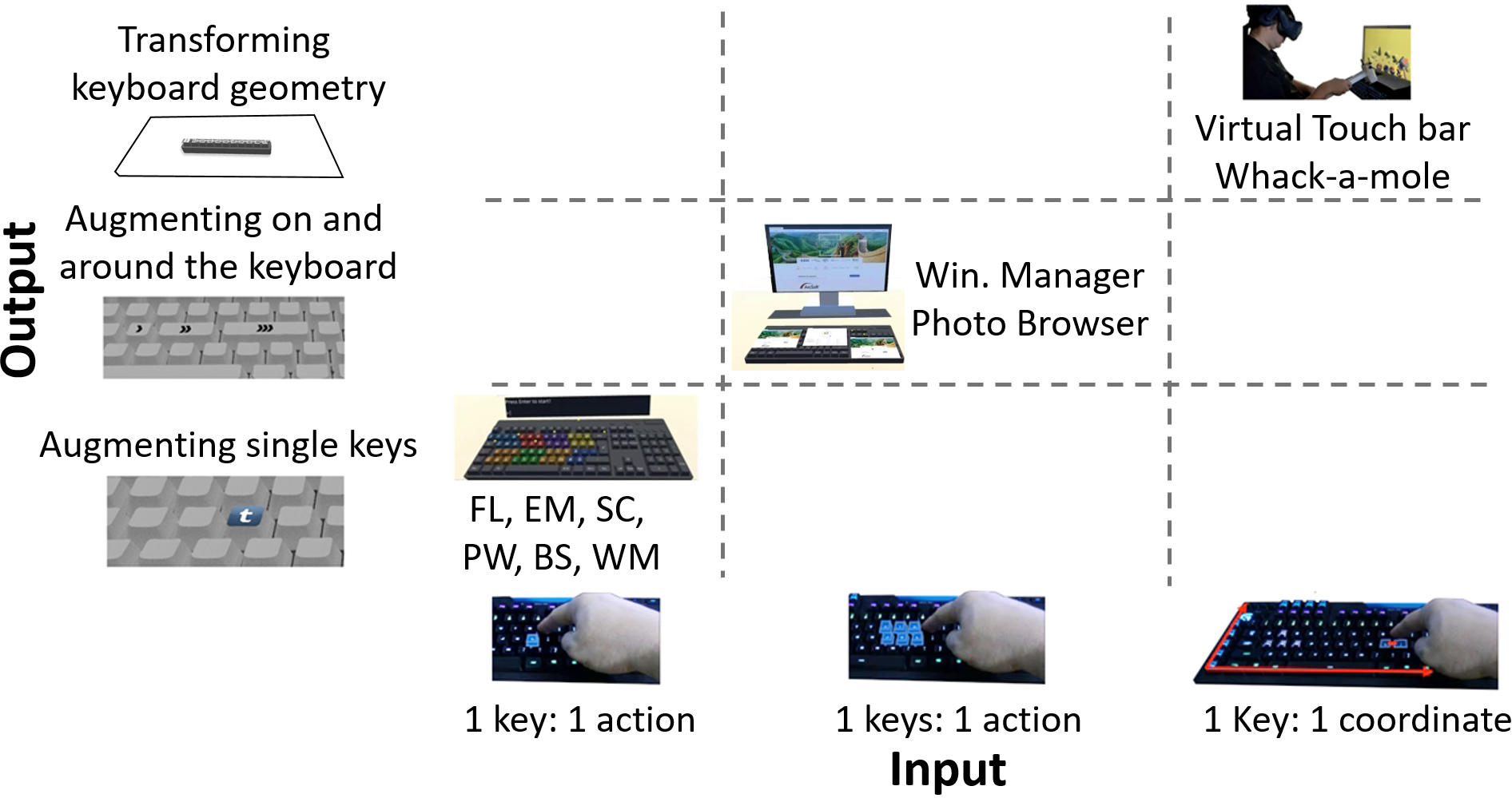}
		\caption{Input-output dimensions of reconfiguring physical keyboards in virtual reality with mapped example applications. The $x$-axis shows input mappings and the $y$-axis shows output mappings. FL: Foreign Languages; EM: Emojis; SC: Special Characters; PW: Secure Password Entry; BS: Browser Shortcuts; WM = Word Macros.}
		\label{fig:designspace2}
		\vspace{-0.5cm}
	\end{figure}

	In contrast, outside the domain of VR, the idea of repurposing the physical keyboard for more than plain text entry has sparked several research projects regarding how to extend sensing capabilities of keyboards (e.g.~\cite{Taylor:2014:T9B:2556288.2557030}) or on how to change the visual appearance of keys (e.g.~\cite{block2010touch}). We see great potential of taking the idea of reconfiguring physical keyboards and apply it into the domain of VR. Hence, this paper examines how to reconfigure the visual representation and functionality of the keyboard itself. In VR, the keyboard can easily display different graphical elements on its keys and change the number, shape and size of the keys, and enable completely new functionally, such as 1D sliders and image maps. To guide the exploration of different keyboard usages, we introduce \emph{input} and \emph{output mappings}, see Figure \ref{fig:designspace2}.
	

	Our contributions in this paper are as follows.
	We transplant the idea of reconfiguring a physical keyboard to virtual reality and explore a set of possible input and output mappings by designing and implementing nine VR applications that we evaluate in a study. We focus on widely available standard physical keyboards, and, hence, do not investigate the possible input space on and above the keyboard \cite{marquardt2011continuous}. We investigate the feasibility of the applications in a user study ($N$=20) and find that they are usable in VR. For two applications (password entry, virtual touch bar) investigated in depth, we find the following: 1) Our results indicate that for password entry shuffling keys in a local region is sufficient for secure password entry; 2) For a selection-based task on a virtual touch bar, changing the visual representation of the keyboard does not affect task performance, however, it has side effects on user's perceived usability and on accidental collisions with the keyboard. We then discuss the limitations and possibilities of remapping the input and output characteristics of physical keyboards in VR based on our empirical findings and analysis and suggest future research directions in this area.


	
	\section{Related Work}
	
	Our work touches on the areas of physical keyboards in VR, augmenting physical keyboards outside of VR,  passive haptics, and security in mixed reality.

	
	
	
	

	\subsection{Physical Keyboards for VR}
	
	While various modalities can be utilized for text entry in Virtual Reality (for a survey see \cite{dube2019survey}), recent research has focused on the feasibility of typing on physical full-sized keyboards in VR.
	An obvious problem is the lack of visual feedback. Without visual feedback users' typing performance degraded substantially. However, by blending video of the user's hands into VR the adverse performance differential significantly reduced \cite{mcgill2015dose}.
	Fundamentally, there are three solution strategies for supporting keyboards in VR. First, by providing complete visual feedback by blending the user's hands into VR. Second, by decoding (auto-correcting) the user's typing to compensate for noise induced by the lack of feedback. Third, by investigating hybrid approaches, such as minimal visual feedback, which may or may not require a decoder to compensate for any noise induced by the method.
	

	Walker et al.~\cite{walker2016decoder, walker2017efficient} investigated typing on a physical keyboard with no visual feedback. They found that the character error rate (CER) was unacceptably high but could be reduced to an average 3.5\% CER using an auto-correcting decoder.
	McGill et al.~\cite{mcgill2015dose} investigated typing on a physical keyboard in Augmented Virtuality \cite{milgram1994taxonomy}. Specifically, they compared a full keyboard view in reality with a no keyboard condition, a partial and full blending condition. For the blending conditions the authors added a camera view of a partial or full scene into the virtual environment as a billboard without depth cues. They found, that providing a view of the keyboard (partial or full blending) has a positive effect on typing performance. Their implementation is restricted to typing with a monoscopic view of the keyboard and hands and the visualization of hand movements is bound by the update rate of the employed camera (typically 30 Hz).  Similar approaches have been proposed \cite{gray2018facilitating} and commercialized  \cite{bovet2018using, VIVEKB}. 
	Lin et al. \cite{lin2017visualizing} investigated the effects of different keyboard representations on user performance and preference for typing in VR but did not study different hand representations in depth.
	
	
	
	Grubert et al. \cite{grubert2018text, grubert2018hand}  investigated the performance of physical and touch keyboards and  physical/virtual co-location for VR text entry. They proposed to use minimalistic fingertip rendering as hand representation and indicated that this representation is as efficient as a video-see through of the user's physical hands \cite{grubert2018hand}. Subsequently, similar studies have investigated further hand representations, such as semi-transparent hand models \cite{knierim2018physical}. Besides, optical outside-in tracking systems with sub-millimeter accuracy such as Optitrack Prime series (e.g., in \cite{grubert2018text, grubert2018hand, knierim2018physical}), also commodity tracking devices such as the Leap Motion have been utilized \cite{hoppe2018qvrty} for hand tracking on physical keyboards. Our work extends these previous works by investigating how different input and output mapping on physical keyboards can be utilized in VR for tasks beyond text entry. 

	\subsection{Augmenting Physical Keyboards}
	
	There have been several approaches in extending the basic input capabilities of physical keyboard beyond individual button presses. Specifically, input on, above and around the keyboard surface have been proposed using acoustic \cite{kurosawa2013keyboard, kato2010surfboard}, pressure \cite{dietz2009practical, zagler2003fasty, loy2005development}, proximity  \cite{taylor2014type}, capacitive sensors \cite{fallot2006touch, habib2009dgts, tung2015flickboard, rekimoto2003presense, shi2018gestakey, block2010touch}, cameras \cite{wilson2006robust, kim2014retrodepth, ramos2016keyboard}, body-worn orientation sensors \cite{buschek2018extending} or even unmodified physical keyboards \cite{lee2013multidirectional, zhang2014gestkeyboard}. Besides sensing, actuation of keys has also been explored \cite{bailly2013metamorphe}.
	
	Embedding capacitive sensing into keyboards has been studied by various researchers. It lends itself  to detect finger events on and slightly above keys and can be integrated into mass-manufacturing processes. Rekimoto et al. \cite{rekimoto2003presense} investigated capacitive sensing on a keypad, but not a full keyboard. Habib et al. \cite{habib2009dgts} and Tung et al. \cite{tung2015flickboard} proposed to use capacitive sensing embedded into a full physical keyboard to allow touchpad operation on the keyboard surface. Tung et al. \cite{tung2015flickboard} developed a classifier to automatically distinguish between text entry and touchpad mode on the keyboard. Shi et al. developed microgestures on capacitive sensing keys \cite{shi2017gestakey, shi2018gestakey}. Similarly, Zheng et al. \cite{zheng2016finger, zheng2018fingerarc} explored various interaction mappings for finger and hand postures. Sekimoro et al. focused on exploring gestural interactions on the space bar \cite{sekimori2018ex}. While we acknowledge the power of extending the sensing capabilities of physical keyboards, within this work we concentrate on the possibilities of unmodified standard physical keyboards. 
	
	Extending the idea of  LCD/OLED-programmable keyboards \cite{LCDKeysHistory, RDS, OPTIMUS}, Block et al. extended the output capabilities of touch-sensitive, capacitive sensing keyboard by using a top-mounted projector \cite{block2010touch, gellersen2012novel}. Block et al. demonstrate several applications that also inspired our work but they did not evaluate them in a user study. Several commercial products have also augmented physical keyboards with additional, partly interactive, displays (e.g., Apple Touch Bar \cite{ATB}, Logitech G19 \cite{LTG19}, Razer Death-Stalker Ultimate \cite{RDS}). Our work builds on these ideas of changing the physical representations of individual keys or keyboard areas, iterates on them in VR and extends them to change the visual representation of the whole keyboard.
	
	\subsection{Usable Security in Mixed Reality}
	
	The idea of usable security \cite{sasse2005usable} has been explored within the domain of Augmented and Virtual Reality \cite{roesner2014security, lebeck2017securing, george2017seamless, alsulaiman2006novel}. De Guzman et al. provide a recent overview of this growing field \cite{de2018security}. The closest work to ours is by Maiti et al. \cite{maiti2017preventing} who explore the use of randomized keyboard layouts on physical keyboards using an optical see-through display. Our work is inspired by this idea, translates it and evaluates it in VR.

	


	\begin{figure*}[ht]
		\centering
		\includegraphics[width=2\columnwidth]{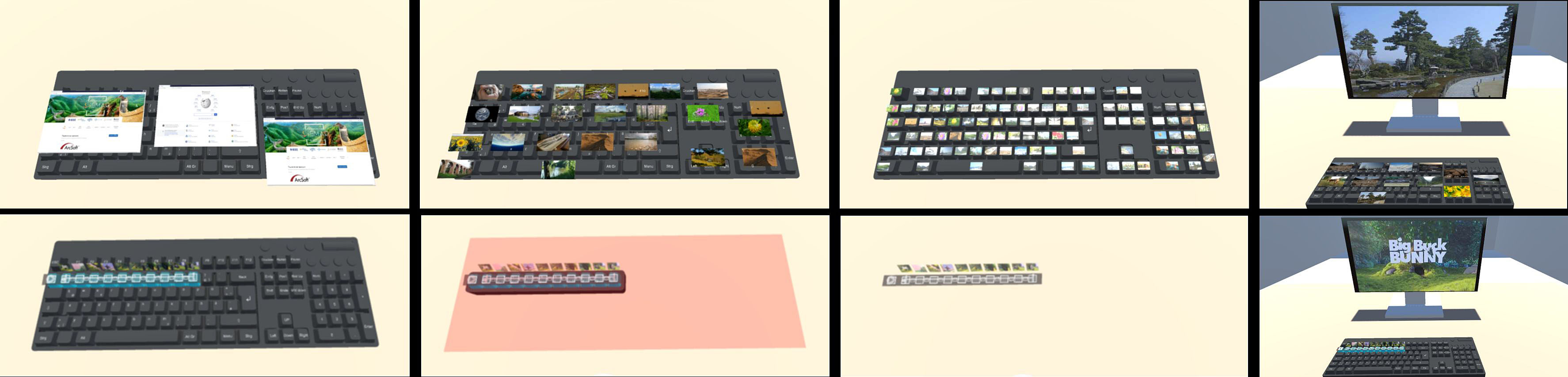}
		\caption{Top row from left to right: Window manager with three virtual buttons, photo browser with 24 and 104 images, view on virtual monitor with selected image. Bottom row from left to right: Virtual touchbar for controlling the timeline of a movie with highlighted keys, as one row keyboard plus red bounding box, without any visualized key, view on the virtual monitor with selected frame.}
		\label{fig:appskeyboard}
	\end{figure*}
	
	\subsection{Passive Haptics for VR}
	
	Passive haptics uses existing props to provide haptic feedback for virtual objects \cite{lindeman1999hand}. The user does not need to wear any physical object, but rather comes into contact with it when interacting with a virtual object. Hinckley et al. \cite{hinckley94passive} seminal work was followed by \cite{burns07machbeth}, \cite{conti05spanning}, \cite{Fitzmaurice95spanning} and may be used to simulate object ranging from hand held ones to walls and terrain [\cite{hughes05mixed}, \cite{low01life}. 
	
	Using existing props raises several challenges addressed by past works. First, a need of a shape similarity between approximating  physical object to an virtual one limits the variation of virtual objects that can be represented by a given physical object. Cheng et. al \cite{cheng17Sparse} used redirection of the users' hand to map a variety of virtual geometries to a fixed physical proxy and measure a difference between real and virtual geometries. Zhao and Follmer \cite{zhao18functional} suggested a parametric optimization to calculate needed retargeting. In an effort to create haptic proxies that are close to the virtual object shapes, researchers printed specific physical props \cite{hinckley94passive}\cite{Kruszynski09mixed}. Shape display devices, are active shape changing devices that try to mimic the virtual object geometry, such as LineForm rendering curves [Nakaguki 15], actuated pins arrays (\cite{follmer13inform}, \cite{kontarinis95tactile}), and constructive blocks \cite{roudaut16chanibals}, \cite{roudaut14chanibals}. Teather et al. \cite{teather2009evaluating} evaluated the effects of co-location of control and display space in fishtank VR and found only subtle
	effects of co-location in an object movement task using a tracked stylus. Similarly, further research on visuo-motor co-location on 3D spatial tasks resulted in inconclusive results, not indicating statistically significant differences \cite{fluet2012effects, fu2012effect}.
	
	In this paper, we do look at an existing physical object: a keyboard. While every surface that is used for touch is represented at the exact location in the virtual space, we do look at ways in which we can modify the keyboard in the virtual world – highlighting new functionalities and obfuscating unused parts of the keyboard.
	
	A second challenge is the need to track the position of the physical proxy in space. Corsten et al. \cite{corsten13functional} provide the ability to recognize and track these physical props using Kinect. Since we use a dedicated device – a keyboard, we can mark it to ease tracking by the virtual reality system. We may use active trackers such as Vive Tracker \cite{VIVEKB} or passive visual trackers, as used in this paper.
	
	Finally, there is a challenge to populate the user’s physical environment with a range of passive objects that may be used as proxies for a slew of virtual objects. Hettiarachchi et al. opportunistically repurpose physical objects to provide haptic responses to specific widget-like digital controls, such as sliders and buttons. Annexing reality \cite{hettiarachchi16annexing} analyzes the environment and opportunistically assigns objects as passive proxies. Azmandian et al. \cite{Azmandian:2016:HRD} allow the use of one such passive proxy for several virtual ones in the same virtual space, yet the problem of finding a close enough proxy in the environment is required. In this work an existent of a physical keyboard in the user’s vicinity is assumed.

	\section{Reconfiguring Physical Keyboards in VR}

	The fundamental dimensions that directly tie into physical keyboard reconfiguration are its \emph{input mapping} and \emph{output mapping}, see Figure \ref{fig:designspace2}. The first input mapping is a direct mapping of a physical key to an action, which effectively results in the key being overloaded. Examples include any method that reconfigures key labels and their corresponding actions, such as displaying emoticons or special characters on the keys and then outputting the corresponding emoticon or symbol. The second input mapping is to map multiple physical keys to the same single action. This allows a section of the keyboard to correspond to the same user interface element. For example, a user interface element such as a photo, where parts of the photo are mapped to different keys but selecting any such key trigger the same action, such as selecting the photo. This can be convenient when there is not enough space on a single key to display a photo or an application that could be selected, or when we want to make a set of keys stand out as a larger key. The third input mapping uses an area of the keyboard as a map where each single physical key is mapped to a single coordinate in the map.
	
\begin{figure}[ht]
		\centering
		\includegraphics[width=0.9\columnwidth]{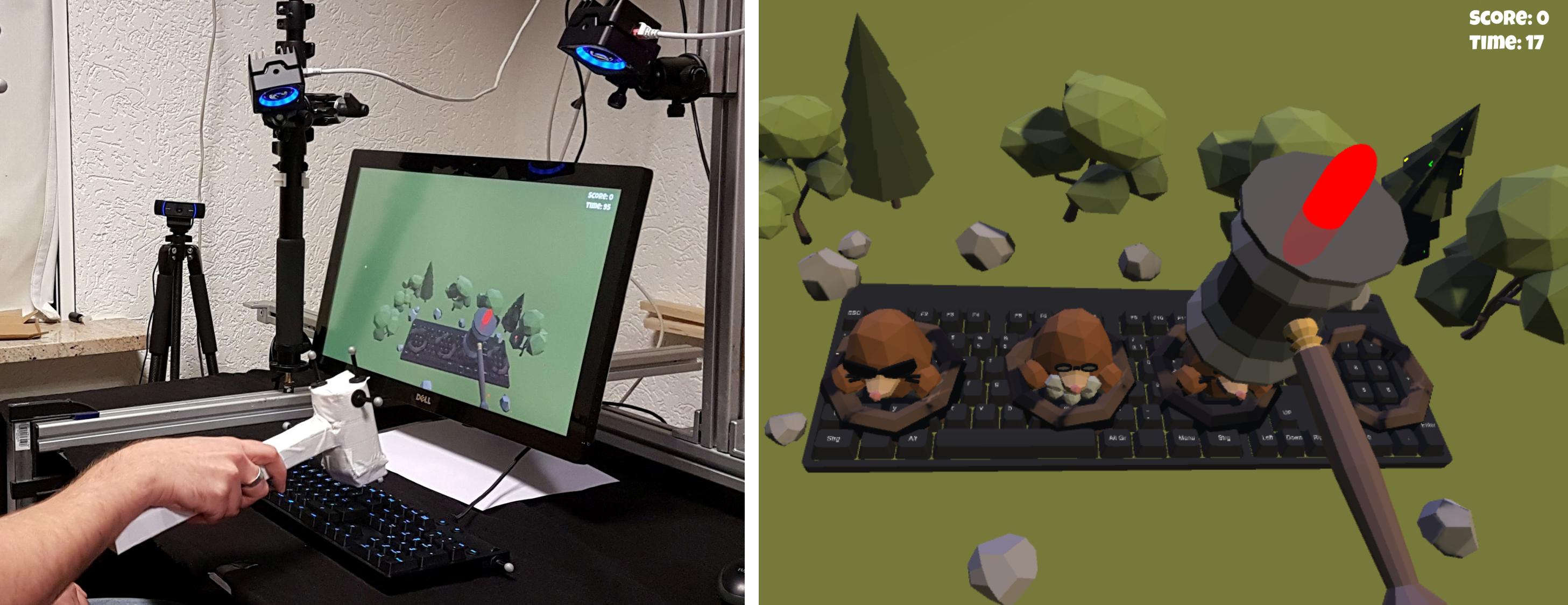}
		\caption{Virtual Whack-A-Mole. Left: User hammering on the keyboard with a spatially-tracked prop. Right: view on the virtual scene.}
		\label{fig:mole01}
	\end{figure}

	We consider three characteristics emerging from the output dimension. First, output can be revealed by augmenting a single key in VR. This allows individual keys to take on specific visual functions, such as for example each key representing a distinct emoticon or special character. Second, output can be generated by augmenting the visual display on and around the keyboard. For example, several keys (and the spaces in between them) can be mapped to displaying a single user interface element, such as a photo. Alternatively, the entire keyboard and its surrounding can be reappropriated as an integrated environment with its individual interface elements mapped to a single or multiple keys. An example of this is the whack-a-mole game shown in Figure \ref{fig:mole01}. Third, the keyboard geometry itself can be transformed into a user interface control with different characteristics, such as an output design of part of the keyboard that resembles a continuous 1D slider. While selection on the slider is discrete due to the input modality still consisting of individual physical keys, the output modality appears continuous to the user.

	\subsection{Applications}
	\label{sec:applications}
	
	
	Clearly it is impossible to exhaustively explore the rich input and output mappings that are realizable when allowing physical keyboards to be reconfigured for VR. Instead, we probed the input-output dimensions by selecting a diverse set of VR applications within the input and output mappings shown in Figure \ref{fig:designspace2}. The applications were selected by considering several factors that collectively result in a diverse set of viable and interesting VR applications: the type of computer tasks people are engaging with today and how they can be improved upon or supported in VR, the degree of input keyboard reconfigurability required, the level of visual output design inside and around the keyboard, the task complexity, and objective of the user (such as office work, rapid access to functionality, window multitask support and entertainment).
	
	
	\subsubsection{Emojis}
	Emojis, i.e., Unicode graphic symbols, used as a shorthand to express concepts and ideas are increasingly used in mobile scenarios \cite{novak2015sentiment}. To support efficient emoji entry using a physical keyboard, we enabled emoji entry through pressing a modifier key (e.g., CTRL) that switches the normal keyboard mapping to an emoji set, see Figure \ref{fig:appskeys}, top row, left. To support an increasing number of emojis, the emoji set can be changed by pressing the modifier key and the page up / down buttons. We will refer to this application as \textsc{Emojis} in the remainder of this paper.
	
	\subsubsection{Languages and Special Characters}
	For those who write in multiple languages, the respective character mappings can  be visualized on the virtual keyboard, see Figure \ref{fig:appskeys}, bottom row, left, for Cyrillic characters. Supporting multiple languages on keyboard is becoming very popular in mobile scenarios with touch keyboards. We implemented language mappings for Arabic, Cyrillic, Greek, Hindi and Japanese. Languages can be switched by pressing a modifier key and the page up / down buttons.
	
	Further, special characters (such as umlauts) can be entered by first pressing a modifier key (in this case, ALT) and the respective letter (e.g., ''a'')  using one hand. Then the neighboring keys are highlighted in green and will temporarily show the available umlauts, see Figure \ref{fig:appskeys}, top row, middle. The umlaut can then be selected by pressing the respective green key with a third finger (e.g., using the second hand). We will refer to this application as \textsc{Languages} in the remainder of this paper.
	
	\subsubsection{Application Shortcuts}
	Keyboard shortcuts for applications have the benefit of triggering actions fast but at the cost of memorizing these shortcuts, which can become challenging with shortcuts reaching dozens to hundreds in applications such as Adobe Photosop or Microsoft Office \cite{omanson2010comparison, lane2005hidden}.
	
	In order to support the keyboard shortcut discoverability, we implemented shortcuts for a virtual Web-browser (based on Zenfulcrum\footnote{https://zenfulcrum.com/, last accessed March 19, 2019.}), see Figure \ref{fig:appskeys}, bottom row, middle. Specifically, we implemented navigation shortcuts (back, forth, home), refresh, cancel and access to 10 bookmarked webpages. We will refer to this application as \textsc{BrowserShortcuts} in the remainder of this paper.
	
	\subsubsection{Virtual Text Processing Macros}
	While keyboard shortcuts are often reflecting a predefined set of actions, new application actions can also be defined. To this end, we defined macros in Microsoft Word (insert signature, insert sender's address, insert image) and mapped them to individual keys (again triggered by a modifier key) on the virtual keyboard. Inclusion of desktop applications such as Word can be achieved by utilizing a virtual desktop mirror\footnote{https://github.com/Clodo76/vr-desktop-mirror, last accessed March 19, 2019}. We will refer to this application as \textsc{WordMacros} in the remainder of this paper.
	
	\subsubsection{Window Manager}
	Various schemes for switching open applications (e.g., using ALT-TAB on Windows) have been proposed for Window managers. In VR, it is possible to visualize open applications directly on the keyboard, see Figure \ref{fig:appskeyboard}, top row, left. Alternative selection (to alt-tabbing or mouse-based selection), can be achieved by pressing one key in the area below the visualized open window. We will refer to this application as \textsc{WindowManager} in the remainder of this paper.
	
	\subsubsection{Photo Browser}
	Similarly, to switching open windows, browsing and selecting photos can be achieved. In a virtual photo browsing application a set of available photos is mapped to the keyboard, see Figure \ref{fig:appskeyboard}, top row, second and third from left to right. The number of photos to be visualized at once is limited by the number of physical keys on the keyboard (e.g., 104 or 105). Similarly, to the emoji application, further photos can be shown when buttons (such as page up / down) are reserved for toggling between photo sets. We will refer to this application as \textsc{PhotoBrowser} in the remainder of this paper.
	
	
	\subsubsection{Whack-a-Mole}
	Besides productivity tasks, the keyboard can also be utilized for leisure applications. We implemented a  whack-a-mole game by turning the keyboard into a virtual playground. Moles are digging there way through the ground and have to be whacked with a virtual hammer. To this end, we utilized a physical hammer prop, that is spatially tracked, see Figure \ref{fig:mole01}, but also implemented a version that does not rely on a physical prop (basically triggering a hammer movement when a keyboard area is hit by the user's hand). We will refer to this application as \textsc{WhackAMole} in the remainder of this paper.
	
	The Whack-a-Mole is just one example of usage of the keyboard surface as an image map, and enabling a very quick accessing a map point by pressing a key.  Although this application is a game and uses a hammer, we could also imagine dynamic modifications of visuals on or over the keyboard used for learning applications. With \textsc{WhackAMole}, we display moles and ask the user the react quickly. As an example, this concept could translate to typing applications, where we could highlight which key(s) should be tapped with which finger(s) of which hand and measure the user's reaction time. In addition, for beginner typists, we could also display over the keyboard a semi-transparent 3D model of a hand to explain how the hand should approach the keyboard for each key as the user is typing.
	
	\subsubsection{Secure Password Entry}
	\label{}
\begin{figure}[ht]
		\centering
		\includegraphics[width=0.7\columnwidth]{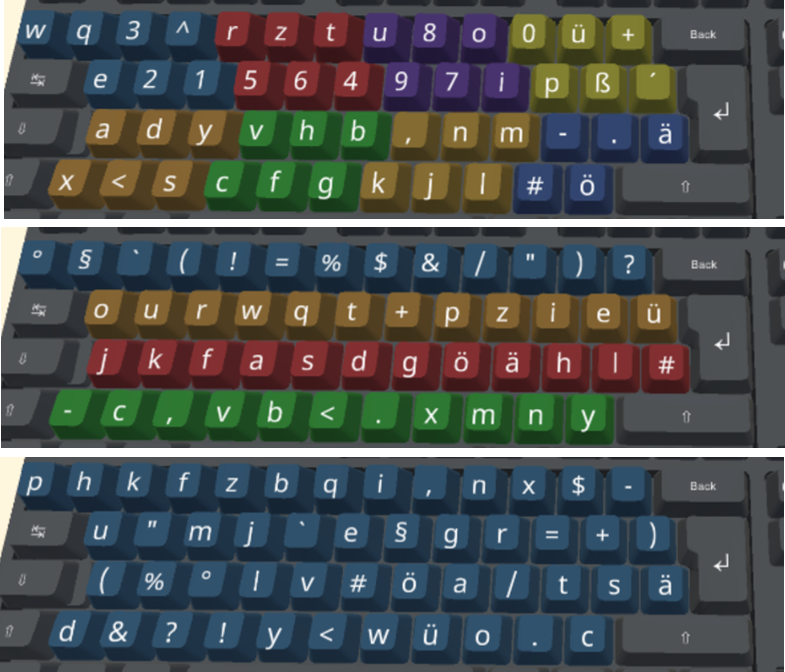}
		\caption{Conditions in the password entry experiment. From top to bottom: \textsc{RegionsShuffle}, \textsc{RowShuffle}, \textsc{FullShuffle}.}
		\label{fig:condpw}
	\end{figure}

	When engaging with a physical keyboard while wearing a immersive HMD, users can become unaware of their surroundings \cite{mcgill2015dose}. This can become critical, when users enter sensitive data such as passwords. To support users in minimizing the risks of shoulder surfer attacks \cite{eiband2017understanding}, we implemented three different key randomization strategies, see Figure \ref{fig:condpw}, which offer varying trade-offs between guessability of the pressed key by a shoulder surfer and discoverability of the key by the VR user. In \textsc{RegionsShuffle}, keys are randomized in the local region of the key. In \textsc{RowShuffle}, keys are randomly assigned along the original row. In \textsc{FullShuffle}, keys are randomly assigned across the keyboard. Please note, that \textsc{RowShuffle} and \textsc{FullShuffle} have been proposed in the domain of AR text entry \cite{maiti2017preventing}. As a first-order approximation, we estimate the probability $p$ of an observer correctly guessing a password of length $n$ as:
	
	\begin{equation}
	\label{mod1}
	p = k^{-n},
	\end{equation}
	where $k$ is the number of keys that are shuffled. This formula assumes 1) the password is truly random; 2) an observer has perfect ability to always correctly infer the key location of any key press of the user; and 3) the observer knows the shuffling system (Kerckhoffs's principle).
	
	While passwords in practice are not truly random, we can use the above estimation to create an illustration of the theoretical trade-off between the effort incurred by the user in searching for the keys on a shuffled keyboard against the probability of an observer inferring the password by observing the user's hand movement.
	The time $T$ it takes for a user to type a password of length $n$ with $k$ keys shuffled can be estimated as:
	\begin{equation}
	\label{mod2}
	T = n \cdot KT + (1-\alpha)(n-1)(k \cdot DT) + k \cdot DT,
	\end{equation}
	where ${KT}$ is average time to move to a key and press it, ${DT}$ is average decision time, that is, the time it takes a user to look at a key and decide whether it is the intended key or not, and $\alpha \in [0,1]$ is a parameter specifying the user's ability to memorize the location of the shuffled keys after an initial scan ($\alpha = 0$: no memory; $\alpha = 1$: perfect memory).
	
	Equations \ref{mod1} and \ref{mod2} form a system that captures the trade-off in time required by the user to type the password compared to the probability of an observer inferring the password. ${KT}$ and ${DT}$ are empirical parameters that can be held constant while $n$ and $k$ are controllable parameters of the system and $\alpha$ is an uncontrollable parameter that varies according to the individual user's ability to quickly remember the shuffled layout (and depending on how often the system changes the key layout). Thus  setting ${KT}$ and ${DT}$ to appropriate values and holding $n$, $k$ and $\alpha$ constant reveals the trade-off between time incurred to type a password and the probability of an observer inferring the password as a function of password length, the number of shuffled keys and the ability of the user to memorize the shuffled layout.
	
	For example, assume the desired probability of an observer at any given instance is able to infer the password is set to one in a million and password length is set to 8. Solving for the integer $k$ in Equation \ref{mod1} results in $k = \lceil p^{-\frac{1}{n}}\rceil$ and thus the number of keys that should be shuffled is 6. Using nominal values for typing a letter of random text and deciding whether an individual key is the intended one \cite{card2018psychology} provides estimations of ${KT} = 0.5$ s and ${DT} = 0.24$ s. At this operating point, Equation \ref{mod2} then provides a time estimate of approximately 5.5 s for typing the password if the user has perfect memory of the reshuffled layout after the first scan and 15.5 s if the user has no memory. As Equation \ref{mod2} is a linear combination, the user's ability to memorize the layout after a scan will linearly affect the time prediction at any given point between the two extremes. We emphasize that we do not make an attempt to introduce an accurate model but merely wish to illustrate that a first-order approximation is sufficient for making the inherent trade-offs of the design parameters explicit to the designer.
	
	\subsubsection{Virtual Touch Bar}
	Inspired from the ideas of LCD/OLED-programmable keyboards \cite{LCDKeysHistory, RDS, OPTIMUS} and, in recent years, keyboards augmented with additional displays (e.g., Apple Touch Bar \cite{ATB}) we explore how ordinary physical keyboards can be turned into quasi touch bar keyboards in VR. While a physical keyboard lacks the input resolution of an actual capacitive-sensing touch bar, a physical keyboard in VR has the potential benefit of 3D content visualization. Further, the visualizations are not limited to solely augment the keyboard.
	
	In our prototype, We implemented an application for controlling the seek bar of a virtual video player. In order to increase the input resolution of the virtual touch bar keys (number keys was between 1 and 10), users could press two adjacent buttons to select an intermediate frame, e.g., if pressing key 1 would jump to second 10 in a 100 second film and key 2 would jump to second 20, then pressing keys 1 and 2 simultaneously would jump to second 15. 
	
	As a base visualization, we highlighted relevant keys for interaction, as shown in Figure \ref{fig:appskeyboard}, bottom row, left. We will call this \textsc{VTHighlight} in the remainder of the paper.
	
	VR allows us to change the visual representation of individual keys or the whole keyboard. To explore this idea, we implemented a one row keyboard that is complemented by a colored bounding box that indicates the geometric bounds of the whole keyboard, see Figure \ref{fig:appskeyboard}, bottom row, second from the left). The design rationale behind this choice was to allow users to have an accurate visual representation of the physical keys that are relevant for interaction and, at the same time, to give a visual indication of the physical dimensions of the inactive part of the keyboard. We call this implementation \textsc{VTOneRow} in the remainder of the paper.
	
	A further variation is to completely hide the keyboard and replace it by a relevant graphical user interface representation for the task at hand. To this end, we only indicated a 1D slider element at the place of the original keys (see Figure \ref{fig:appskeyboard}, bottom row, third from left). We call this implementation  \textsc{VTInvisible} in the remainder of the paper. Each variant has its own potential benefits and drawbacks. The most apparent design issue is perceived accordance \cite{norman2013design}: will users be able to perceive the given functionality given the visual representation? This is explored further in the user study in the next section.
	
	\section{User Study}
	
	The purpose of our user study was threefold. First, we wanted to get feedback to learn from initial user reactions. To this end, we followed the approach by Chen et al. of demonstrating individual experiences to users instead of carrying out task-based evaluations \cite{chen2014duet}. To this end, we demonstrated the following applications described in Section \ref{sec:applications}: \textsc{WhackAMole}, \textsc{PhotoBrowser}, \textsc{WindowManager}, \textsc{WordMacros}, \textsc{BrowserShortcuts}, \textsc{Languages}, \textsc{Emojis} and asked the participants to engage with each application. Second, we wanted to understand how physical keyboards can be utilized to support usable security in the context of shoulder surfing attacks in VR. Similar studies have been conducted in AR \cite{maiti2017preventing}. However, it is unclear how password entry using shuffled keys translates to VR due to the different output media (e.g., AR glasses project the keys into a different depth layer than the physical keys). Further, and in contrast to previous work, our focus was to better understand the relation between objective and perceived security as well as the trade-off between perceived security and text entry performance. Third, we wanted to investigate the effects of changing the visual representation of a physical keyboard on user experience and performance. To this end, we employed a selection-based task in context of the virtual touch bar app.

	\begin{figure}[ht]
		\centering
		\includegraphics[width=0.7\columnwidth]{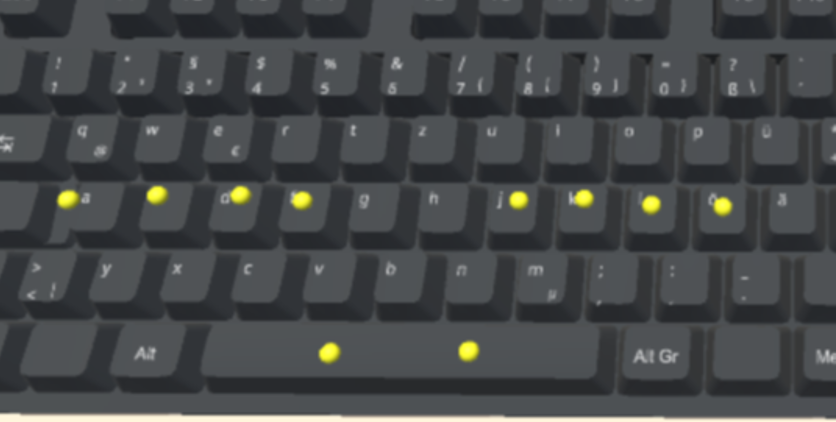}
		\caption{Spheres used as fingertip visualization during the experiment.}
		\label{fig:hands}
	\end{figure}

	\subsection{Participants}
	
	We recruited 20 participants from a university campus with diverse study backgrounds. 
	All participants were familiar with QWERTZ desktop keyboard typing. 
	From the 20 participants (5 female, 15 male, mean age 27.8 years, $sd = 3.8$, mean height 176.3 cm, $sd = 8.6$), 8 indicated to have never used a VR HMD before, four participants once, four participants rarely but more than once, two participants occasionally and two participant to wear it very frequently. Four participants indicated to not play video games, one once, two rarely, 6 occasionally, four frequently and three very frequently. Seven participants indicated to be highly efficient in typing on a physical keyboard and 12 to write with medium efficiency on a physical keyboard and one to write with low efficiency on a physical keyboard (we caution against over-interpreting these self-assessed performance indications). Nine participants wore contact lenses or glasses. Two volunteers have participated in other VR typing experiments before.
	
	\subsection{Apparatus and Materials}
	
\begin{figure}[ht]
		\centering
		\includegraphics[width=0.7\columnwidth]{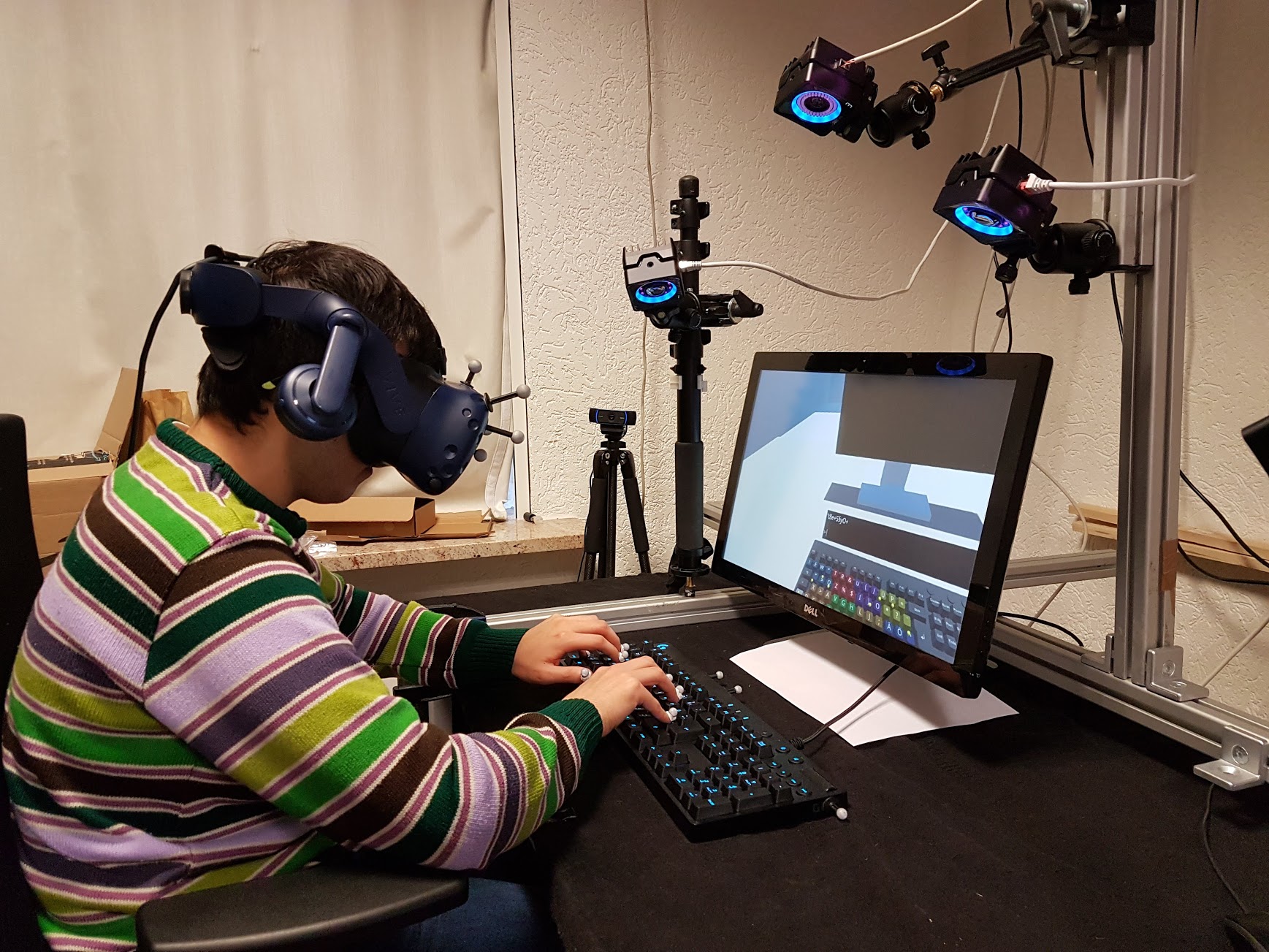}
		\caption{Apparatus of the experiment showing a participant with nail attached retro-reflective markers, the G810 keyboard, a HTC VIVE Pro Headset, the Optitrack Prime 13 tracking system and an external webcam for logging. Please note that the headphones are not attached to the ears.}
		\label{fig:apparatus}
	\end{figure}

	An OptiTrack Prime 13 outside-in tracking system was used for spatial tracking of finger tips, the HMD and the keyboard. The tracking system had a mean spatial accuracy of 0.2 mm. A HTC Vive Pro was used as HMD. As physical keyboard a Logitech G810 was used. The setup is shown in Figure \ref{fig:apparatus}. We utilized a minimalistic finger tip representation as suggested by prior \cite{grubert2018hand} work to indicate the hand position  relative to the keyboard, see Figure \ref{fig:hands}. 
	
	For the passwords we used a set where half of the passwords were popular simple passwords and the other half was split equally to five and 10 character length randomized passwords. The virtual environment was showing the virtual keyboard and a monitor resting on a desk. In line with previous work \cite{grubert2018hand, grubert2018text} the passwords and the entered text were visualized both on the monitor and directly above the keyboard. The system was implemented in Unity 2018.2 and deployed on a PC (Intel Xeon E5-1650 processor, 64 GB RAM, Nvidia GTX 1070 graphics card) running Windows 10.
	
	
	\begin{figure}[ht]
		\centering
		\includegraphics[width=0.7\columnwidth]{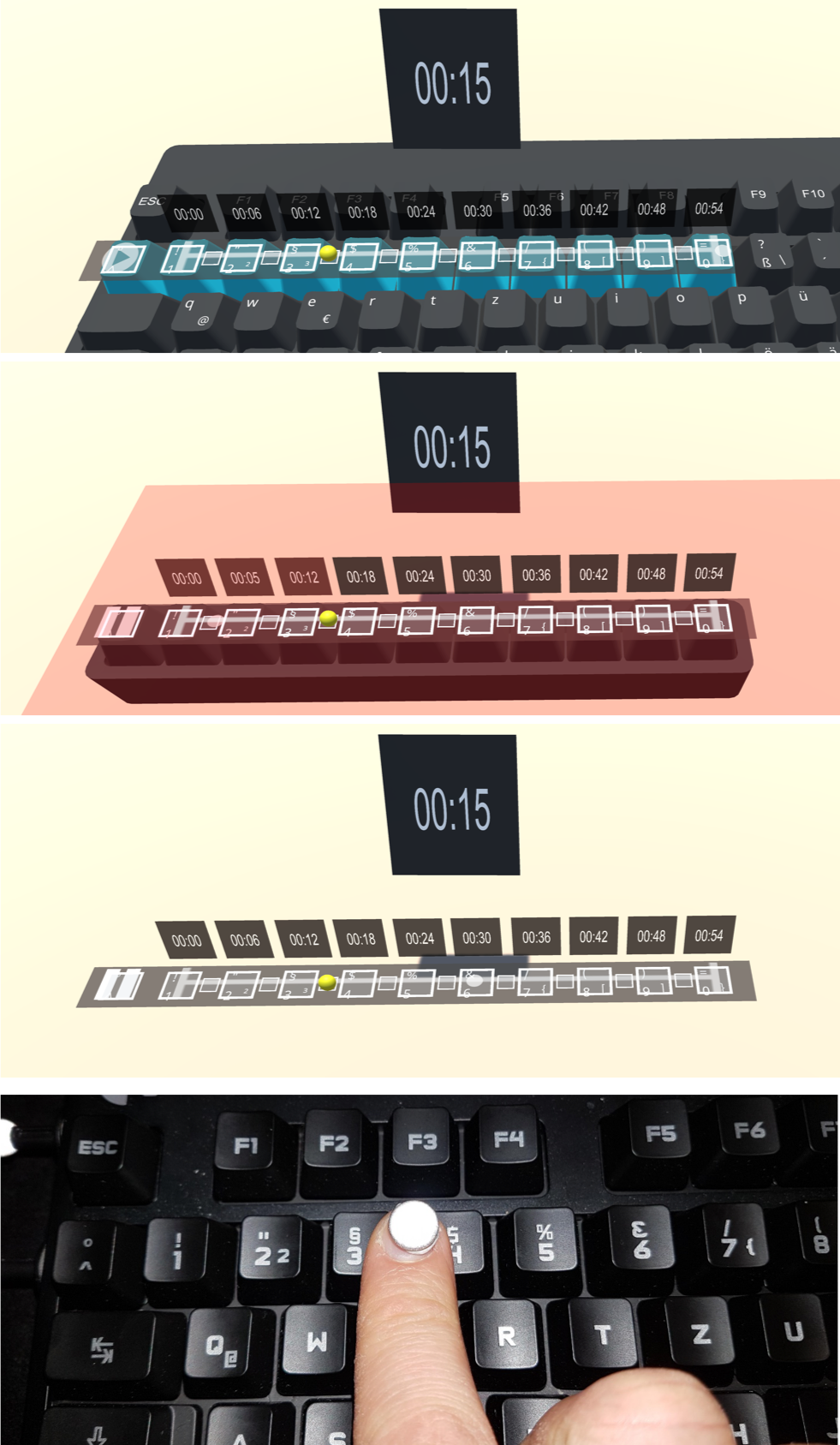}
		\caption{Close-up view on the conditions in the virtual touchbar task. From top to bottom: \textsc{VTHighlight}, \textsc{VTOneRow}, \textsc{VTInvisible}, view on the physical keyboard with a user's finger and attached retro-reflective marker for fingertip tracking. }
		\label{fig:detailsvt}
	\end{figure}

	\subsection{Procedure and Task}
	After welcoming, participants were asked to fill out a demographic questionnaire. Thereafter, we conducted a calibration phase to be able to spatially track the fingers of the participants. Retro-reflective markers were fixated with double sided tape on the participants' fingernails, see Figure \ref{fig:detailsvt}, bottom row. 
	
	The study was divided into three parts with 5-minute breaks in between. The first part gathered user feedback on the seven applications (see below). This part lasted around 30 minutes. After that  evaluation of the password entry (lasting ca. 35 minutes) and virtual touch bar (lasting ca. 20 minutes) applications were carried out alternately. Thus, either of those two applications was executed as the second and the other as the third. After each part, a questionnaire was filled out by the participants to rank the conditions. Then a semi-structured interview followed. On average, the  procedure took 110 minutes. Finally, the participants were compensated with a 10 Euro voucher. Due to the already long study duration we employed a short three-item questionnaire capturing the user experience dimensions of ease of use, utility and enjoyment (in line with similar approaches \cite{lewis1991psychometric, chen2014duet}), instead of utilizing longer questionnaires such as the 10-item system usability scale \cite{brooke1996sus}. 
	
	In the first part, we demonstrated the following seven applications to the participants and asked them to engage with the applications themselves: \textsc{WhackAMole}, \textsc{PhotoBrowser}, \textsc{WindowManager}, \textsc{WordMacros}, \textsc{BrowserShortcuts}, \textsc{Languages}, \textsc{Emojis}.
	We asked for feedback on ease of use, utility and enjoyment for each application after demonstration. 
	The order of the applications was balanced across participants, as good as possible (full permutation was not possible due to the number of applications). 
	The second part alternated with the third part for counterbalancing across the participants.
	To evaluate the password entry application the participants were asked to type passwords for two minutes from a predefined set of the most used passwords\footnote{https://en.wikipedia.org/wiki/List\_of\_the\_most\_common\_passwords, last accessed March 21, 2019}. Two baselines where taken without the HMD and wearing the HMD. For each of the three counterbalanced conditions (\textsc{RegionShuffle}, \textsc{RowShuffle}, \textsc{FullShuffle}) there was a one minute training phase, followed by the 5 minute testing phase. After each condition a questionnaire with questions about ease of use, utility and enjoyment was answered.
	
	The third part, alternating with the second part, was conducted to get insight into changing the visual representation of the keyboard. To this end, we employed the virtual touch bar application, with the three coutnerbalanced conditions (\textsc{VTHighlight}, \textsc{VTOneRow} and \textsc{VTInvisible}), see Figure \ref{fig:detailsvt}. After showing the virtual touch bar in a demo, the participants were asked to use the touch bar with a repeated motion. Starting at a fixated point in the VR (centered 3 cm below the bottom edge of the keyboard), the participants had to move their index finger of the dominant hand to the touch bar and back to the fixated point (basically following the procedure of a Fitts Law task with fixed target size). The fixated point was connected to the coordinates of the VR-keyboard, to guarantee a static distance to the targets. As a baseline 5 timestamps were shown to the participant, that he had to locate in the touch bar. After this the condition was conducted with 25 timestamps. The conditions were counterbalanced across participants.
	

	\subsection{Results}
	Unless otherwise specified, statistical significance tests for performance data (text entry rate, error rate, task completion time) were carried out using general linear model repeated measures analysis of variance with Holm-Bonferroni adjustments for multiple comparisons at an initial significance level $\alpha = 0.05$. We indicate effect sizes whenever feasible ($\eta^2_p$). For subjective feedback, or data that did not follow a normal distribution or could not be transformed to a normal distrubtion using the log-transform, we employed Friedman test with Holm-Bonferroni adjustments for multiple comparisons using Wilcoxon signed-rank tests.
	
	\subsection{Initial user feedback on Applications}
	\label{sec:feedbackapps}
\begin{figure}[ht]
		\centering
		\includegraphics[width=\columnwidth]{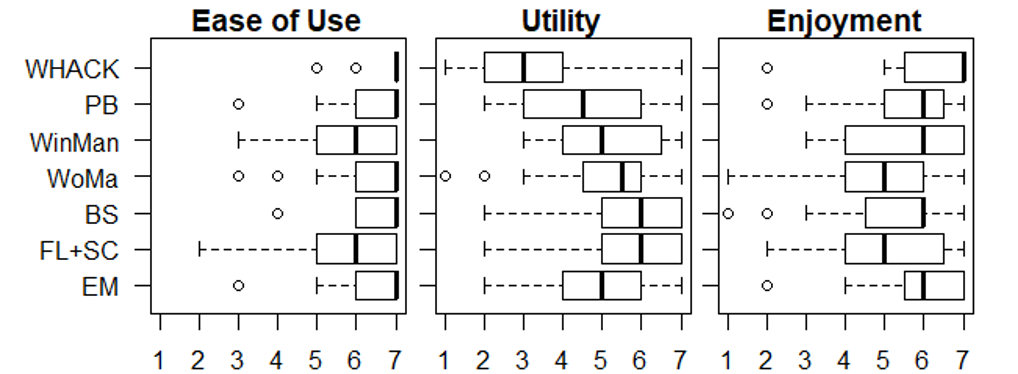}
		\caption{Ease of use ratings, Utility and Enjoyment ratings on a 7-item Likert scale (1: lowest, 7: highest). Abbreviations: WHACK: \textsc{WhackAMole}, PB: \textsc{PhotoBrowser}, WinMan: \textsc{WindowManager}, WoMa: \textsc{WordMacros}, BS: \textsc{BrowserShortcuts}, FL+SC: \textsc{Languages}, EM: \textsc{Emojis}}
		\label{fig:feedbackapps}
	\end{figure}

	Figure \ref{fig:feedbackapps} shows user ratings on seven-item Likert scales for questions on ease of use ("I found the application easy to use"), utility ("I found the application to be useful") and enjoyment ("I had fun interacting with the application"). The figure indicate high ratings for ease of use, 
	varying ratings for utility 
	and for enjoyment. Please note that we did not run null hypothesis significance tests on these ratings as they should serve as a descriptive indication of these user experience dimensions only.

	Users were asked to comment on the individual applications. We followed top-down qualitative coding and structuring procedures to identify benefits and drawbacks of the individual applications \cite{strauss1990basics}. For \textsc{WhackAMole} users mentioned that it was ''fun and totally different experience'', but also that while ''easy to understand'' it is ''not relevant for work'' with one participant stating ''The PC is a tool for me. I prefer things that make work easier''. Regarding both \textsc{PhotoBrowser} and \textsc{WindowManager} users were critical with liking the principal idea but noticing, that when the icons get too small (ca. 4 times the key size), the experience results in ''low usability".
	For \textsc{WordMacros}, opinions were split, with some users mentioning it usefulness and ease of use (e.g., ''I found it useful and easy to use'') and others questioning its utility (''I do not think this is  useful for work''). User's generally appreciated the \textsc{BrowserShortcuts} applications with one mentioning ''Shortcuts are important for productivity. It is not bad to see all of them'' and another one saying ''I need to switch between tabs at work often. It is useful''.
	Regarding foreign languages and special characters in the \textsc{Languages} application, opinions were split for the language mapping according to the cultural background of the users. A user who did not type in multiple languages mentioned explicitly ''This is not useful for me. I do not use other languages'' and one multi-lingual user mentioned ''I find this useful for foreign languages because otherwise it is a lot of work to type special symbols or characters''.
	Regarding special characters opinions were similar, with one user mentioning ''I do not use special characters'' and another one ''I find this useful for formulas and special characters''. However, some users also mentioned the unexpected layout of special characters (in contrast to a simple row on soft keyboards) with one stating: "There is much searching required" to find the needed character. Finally, for \textsc{Emojis}, users generally appreciated the application with one user  ''I would  use this in daily life'' and another one saying ''This is a satisfying way'' (to type emojis) ''even if not in VR''. However, a productivity oriented user also mentioned ''Emojis are unnecessary in daily work''.

	\subsection{Password Entry}
	We report on text entry rate, character error rate, user experience ratings as well as preferences and open comments in the next subsections. In our performance evaluation, we concentrate on the joint set of simple and randomized passwords. While the absolute performance values differ between password sets with respect to text entry speed, the significance  between conditions did not change.
	
	\subsubsection{Entry Rate}
	Entry rate was measured in words-per-minute (wpm), with a word defined as five consecutive characters, including spaces. 
	The entry rate measured in the profiling phase without randomized keys was 21.0 wpm ($sd = 7.91$). 
	The mean entry rate for \textsc{RegionShuffle} was 6.57 wpm ($sd = 1.96$), for \textsc{RowShuffle} it was 6.03 wpm ($sd = 1.63$) and for \textsc{FullShuffle} it was 3.82 wpm ($sd = 1.44$).
	
	An omnibus test revealed significance ($F_{3, 17} = 49.73$, $\eta^2_p = 0.898$, $p < .001$). Holm-Bonferroni adjusted post-hoc testing revealed significant differences between baseline and all randomization layouts (which was to be expected) (adjusted p-values $< 0.001$), between \textsc{FullShuffle} and \textsc{RegionShuffle} (adjusted p-value $< .001$) as well as between \textsc{FullShuffle} and \textsc{RowShuffle} (adjusted p-value $< .001$), but not between \textsc{RegionShuffle} and \textsc{RowShuffle} (adjusted p-value $= 1.00$).
	In other words, \textsc{FullShuffle} lead to significantly reduced text entry speed compared to both \textsc{RegionShuffle} and \textsc{RowShuffle}.
	
	\subsubsection{Error Rate}
	Error rate was measured as character error rate (CER). CER is the minimum number of character-level insertion, deletion and substitution operations required to transform the response text into the stimulus text, divided by the number of characters in the stimulus text. The CER measured in the profiling phase without randomized keys was 3.2\%  ($sd = 3.5$). The CER for \textsc{RegionShuffle} was 3.7\% ($sd = 3.7$), for \textsc{RowShuffle} it was 3.4\% ($sd = 3.4$) and for \textsc{FullShuffle} it was 3.4\% ($sd = 5.3$). An omnibus test revealed no significance ($F_{3, 5} = 0.55$, $\eta^2_p = 0.032$, $p < 0.981$).
	In other words, there where no significant differences in terms of error rate between the conditions.
	
	\subsubsection{User Experience Ratings}
	
\begin{figure}[ht]
		\centering
		\includegraphics[width=0.9\columnwidth]{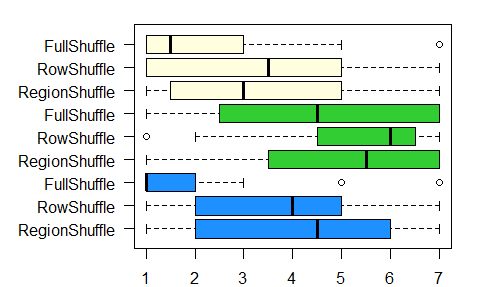}
		\caption{Ease of use (blue), Utility (green) and Enjoyment (yellow) ratings on a 7-item Likert scale (1: lowest, 7: highest). }
		\label{fig:feedbackpw1}
	\end{figure}

	User ratings regarding ease of use, utility and enjoyment (utilizing the same question as in Section \ref{sec:feedbackapps}) are shown in Figure \ref{fig:feedbackpw1}. A Friedman test indicated statistically significant differences between the conditions for ease of use ($\tilde\chi^2 = 23.52$, $p < .001$), and enjoyment ($\tilde\chi^2 = 12.54$, $p = .002$) but not for utility ($\tilde\chi^2 = 0.98$, $p = .61$) .
	
	
	Regarding ease of use, Bonferroni adjusted Wilcoxon signed rank tests indicated pairwise differences between \textsc{Fullshuffle} and \textsc{RegionShuffle} ($Z = -3.55$, $adjusted p < .001$) as well as between \textsc{FullShuffle} and \textsc{RowShuffle} ($Z = -3.43$, $adjusted p = .002$), but not between \textsc{RegionShuffle} and \textsc{RowShuffle} ($Z = -0.722$, $adjusted p = .94$). Regarding enjoyment, Bonferroni adjusted Wilcoxon signed rank tests indicated pairwise differences between \textsc{Fullshuffle} and \textsc{RegionShuffle} ($Z = -2.96$, $adjusted p = .009$) as well as between \textsc{FullShuffle} and \textsc{RowShuffle} ($Z = -2.80$, $adjusted p = .015$), but not between \textsc{RegionShuffle} and \textsc{RowShuffle} ($Z = -0.05$, $adjusted p = .96$).

	We also asked participants to rate the conditions regarding perceived security. The average score on a 7-item Likert scale (1: totally disagree, 7: totally agree) for the statement "I felt protected from shoulder surfers" where 6.10 ($sd = 1.45$) for \textsc{FullShuffle}, 5.60 ($sd = 1.57$) for \textsc{RegionShuffle} and 6.05 ($sd = 1.43$) for \textsc{RowShuffle}. For the statement "I think that the proposed condition makes password entry more secure" the ratings where  6.05 ($sd = 1.28$) for \textsc{FullShuffle}, 5.50 ($sd = 1.32$) for \textsc{RegionShuffle} and 5.60 ($sd = 1.39$) for \textsc{RowShuffle}. While Friedman tests indicated significant differences between conditions, post-hoc comparisons with Bonferroni adjustment failed to indicate pairwise differences. In other words, \textsc{FullShuffle} led to a significant lower ease of use and enjoyment rating compared to both other conditions.
	
	\subsubsection{Preferences and Open Comments}
	\textsc{RegionShuffle} was preferred by 9 participants, \textsc{RowShuffle} by 10 participants and \textsc{FullShuffle} by one participants. Regarding \textsc{RegionShuffle}, four users mentioned ''It is easier to coordinate.'', one  mentioned at this to be the ''most usable'' option and another one commening on the security aspect with saying this option had ''enough shuffling''. For \textsc{RowShuffle} four user mentioned characters are '' easier to find in the row'', one saying ''this is closest to normal use''. However, regarding perceived security two users mentioned a perceived ''low security''. For \textsc{FullShuffle} six users mentioned "it is frustrating" and another six ''It is very time consuming''. Regarding perceived security five users mentioned ''It has the best security''.
	
	\subsection{Virtual Touch Bar}
	We report on task completion time, selection errors, collisions with the keyboard, user experience ratings as well as preferences and open comments in the next subsections. 
	
	\subsubsection{Task Completion Time and Errors}
	The mean task completion time for \textsc{VTHiglight} was 2.15 seconds ($sd = 0.43$), for \textsc{VTOneRow} it was 2.15 seconds as well ($sd = 0.42$) and for \textsc{VTInvisible} it was 2.24 seconds ($sd = 0.39$). An omnibus test did not reveal significance  ($F_{2, 16} = 2.26$, $\eta^2_p = 0.22$, $p = .14$).
	
	The average number of errors (i.e., users pressed a wrong key) was \textsc{VTHiglight} was 0.85 ($sd = 1.39$), for \textsc{VTOneRow} it was 1.55  ($sd = 2.92$) and for \textsc{VTInvisible} it was 1.65 ($sd = 3.94$). A Friedman test did not reveal significance ($\tilde\chi^2 = 3.13$, $p = .21$).
	In other words, all visualizations resulted in comparable performance measures.
	
	\subsubsection{Collisions}
	We observed the number of accidental collisions between the user's hand and the physical keyboard through an external camera and an additional human observer. For \textsc{VTHiglight} the average number of collisions were 0.49 ($sd = 0.82$), for \textsc{VTOneRow} 2.45 ($sd = 2.11$) and for \textsc{VTInvisible} the mean number of collision was 3.65 ($sd = 3.51$).
	
	A Friedman  test indicated statistically significant differences between the conditions ($\tilde\chi^2 = 21.73$, $p < .001$). Bonferroni adjusted Wilcoxon signed rank test indicated pairwise differences between \textsc{VTHiglight} and \textsc{VTOneRow} ($Z = -3.20$, $p = .001$) as well as between \textsc{VTHiglight} and \textsc{VTInvisible} ($Z = -3.740$, $p = .000$), but not between \textsc{VTInvisible} and \textsc{VTOneRow} ($Z = -1.24$, $p = .22$). In other words, showing the full keyboard in\textsc{VTHiglight} led to a significant reduced number of collisions compared to both other visualizations.
	
	\subsubsection{User Experience Ratings}
	
\begin{figure}[ht]
		\centering
		\includegraphics[width=0.9\columnwidth]{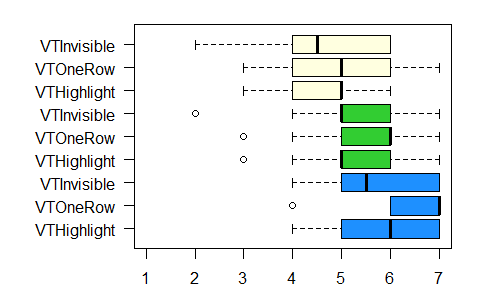}
		\caption{Ease of use (blue), Utility (green) and Enjoyment (yellow) ratings on a 7-item Likert scale (1: lowest, 7: highest). }
		\label{fig:feedbackvt1}
	\end{figure}

	User ratings regarding ease of use, utility and enjoyment (utilizing the same question as in Section \ref{sec:feedbackapps}) are shown in Figure \ref{fig:feedbackvt1}. Friedman tests did not reveal significant differences.
	
	We also asked participants to rate the conditions regarding the following statement "I had the feeling I could hit the targets that I was aiming for" The average score on a 7-item Likert scale (1: totally disagree, 7: totally agree) was 6.30 ($sd = 1.13$) for \textsc{VTHighlight}, 6.15 ($sd = 1.23$) for \textsc{VTOneRow} and 5.53 ($sd = 1.17$) for \textsc{VTInvisible}.
	A Friedman  test indicated statistically significant differences between the conditions ($\tilde\chi^2 = 12.94$, $p = .002$). Bonferroni adjusted Wilcoxon signed rank test indicated pairwise differences between \textsc{VTInvisible} and \textsc{VTHiglight} ($Z = -2.81$, $adjusted p = .015$) as well as between \textsc{VTInvisible} and \textsc{VTOneRow} ($Z = -2.84$, $adjusted p = .015$), but not between \textsc{VTHighlight} and \textsc{VTOneRow} ($Z = -.91$, $adjusted p = 1.0$).

	\subsubsection{Preferences and Open Comments}
	\textsc{VTHighlight} was preferred by 11 participants, \textsc{VTOneRow} by six participants and \textsc{VTInvisible} by three participants.
	
	For \textsc{VTHighlight} five users mentioned ''I have better orientation'' (compared to the other representations), three saying ''I have a better feeling where my fingers are'', two that this representation ''feels most natural'' but also two mentioning that there where ''too many keys" that "are confusing''.
	
	Regarding \textsc{VTOneRow} two participants mentioned that this represenation ''has enough information to navigate'', but also one user mentioning that it ''feels redundant because of the other two options''
	
	Finally, for \textsc{VTInvisible} a user said ''it gives a good overview without useless information''. However, six users mentioned ''It is hard to navigate correctly'' and another one ''It is hard to handle''.

	\section{Discussion}
	The first part of the study indicated that the proposed seven applications \textsc{WhackAMole}, \textsc{PhotoBrowser}, \textsc{WindowManager}, \textsc{WordMacros}, \textsc{BrowserShortcuts}, \textsc{Languages}, \textsc{Emojis} were mostly usable but varied in the utility rating, partly based on the participants' background (e.g., uni- vs. multilingual). However, for both \textsc{PhotoBrowser} and \textsc{WindowManager} participants indicated a reduced usability, when the visualized images approach the size of individual keys. This indicates,  that the input resolution of the keyboard was higher than what was usable for those two applications (which utilized screenshots of applications or websites for visualization). In contrast, for other applications using small symbols (such as \textsc{Emojis}, \textsc{BrowserShortcuts}) no remarks regarding legibility of symbols were made (potentially due to the better discernability of those graphic symbols). 
	
	\subsection{Password Entry}
	The study results for password entry indicated that shuffling the keys over the whole keyboard in condition \textsc{FullShuffle} lead to a significant lower text entry rate (which is to be expected) compared to the other two shuffling schemes and, at the same time, to significantly lower user experience ratings (in terms of ease of use and enjoyment), which is also reflected in the user comments (e.g., ''it is frustrating''). Interestingly, while five users explicitly mentioned that \textsc{FullShuffle} ''has the best security'' the results from the perceived security questions do not fully support this hypothesis due to a lack of statistical significance. Given the current evidence, we would, hence, argue that  regional shuffling of keys seems to be sufficent for future usage as it leads to less user frustration without significantly scarifying perceived security. The first-order model we introduced earlier in this paper would predict a password entry rate for shuffling six keys that lie approximately between 0.5 and 1.5 characters-per-second (cps), which corresponds to between 6 and 18 wpm which roughly correspond to the lower-end to the entry rate observed in the user study.
	
	\subsection{Virtual Touch Bar}
	The experiment indicated that the visual representation of the keyboard can be changed without a significant impact for the specific task at hand. However, we observed a significant higher number of accidental collisions with the physical keyboard for visual representations that do not depict the keyboard in full visual fidelity. In addition, for condition \textsc{VTInvisible}, participants noticed a lower perceived accuracy in aiming for the targets. We actually expected the red bounding box in condition \textsc{VTOneRow} to support users in avoiding accidental collisions, but this is not supported by the experimental results. Hence, further research is needed on appropriate visualizations for non-relevant keyboard areas for a given task at hand.
	

	\subsection{Limitations and Future Work}
	Our work focused on a subset of a possibly large space of input and output mappings. For instance, we didn't explore experiences of augmenting around the keyboard or transform the keyboard when mapping a physical key to an action which we are discussing later in this section. Related, there are additional mappings that can be explored in the future, such as mapping a single key to multiple actions or multiple keys to a coordinate.
	
	Our studies have been based on the physical keyboard Logitech G810 so far, but we see opportunities to explore other types of keyboards including latptop's keyboards. Also, we have been using an external tracker (Optitrack) for our studies, but it would be interesting to see if we could build these type of experiences without relying on fixed trackers in the environment to demonstrate that this approach could really work for mobile users already with today's technology. Also, given the external tracking system, hand and finger actions could be triggered without a keyboard at all. Yet, the act of pressing on a button is subtle and hard to sense by remote sensing, or would require larger gestures. In contrast, using a keyboard button presses can be sensed very accurately and a verification action can be felt. Still, it is valuable to study the effects of using a keyboard vs. on-surface touch for selected tasks in more detail.
	
	In addition, we have results based on the specific experiences we designed. We need to explore more applications to generalize the results. For instance, we studied how to use keys to insert content into Word, but haven't explored yet this type of experience on other types of office productivity applications such as spreadsheet applications. So far, we have used simple visualization in our prototypes, but haven't studied in depth alternate visualizations of physical keyboard reconfiguration in VR.
	
	Moving forward, we are interested in extending the work to combine augmentation of the physical keyboard in VR with augmentation around the keyboard. For instance, window management could be done by displaying all the open applications around the keyboard to allow rapid switching between them. When an application has been selected, the keys of the physical keyboards could be automatically modified accordingly to accommodate the active application. In addition, we are interested in studying if our approach is also working for other type of keyboards. For instance, using a laptop keyboard can be very interesting to explore because we potentially see many people travel with only their laptop and a mobile HMD, a view that is also shared by others \cite{grubert2018office, VRFLIGHT}. Laptop keyboards have different form factor and key design which is likely to open up possibilities for additional design exploration, for instance by allowing gestures by swiping across the physical keys \cite{Zhang:2014:GEG:2556288.2557362}. To this end,  alternative sensing capabilities (such as touch-enabled physical keyboards \cite{otte2019evaluating}) could be employed with our use cases. Further, this work focused on the use of individual fingers when operating the keyboard. Future work should also investigate the use of multi-finger input.
	
	Our work demonstrates the rich design possibilities that open up when reconfiguring physical keyboards for VR. However, this idea can be brought even further by exploring using the mouse in VR along with the keyboard. In fact, the mouse could also be augmented based on the context and the task of the user. We could also consider using a touch mouse, which again will open up additional input mapping possibilities.  Related, touchpads, which are embedded in most laptops, can also be compelling to augment in VR because it is possible to get a very precise coordinate from it and then use this and other input information to dynamically display information on it and modify it's role, such as swapping between acting as a small touchscreen and acting as an indirect pointing device. In some contexts, we could even just display the touchpad and make the keyboard disappear.
	Also, the proposed applications could be transferred to Augmented Reality and explored further. One technical difference in AR is the view on the user’s physical hands. If the view of the physical hands should be adopted (e.g., for showing minimalistic finger representations as in this paper), there is a need to generate a mask of the hands that enables their display or hiding. One option for video see-through AR systems could be to use chroma keying. Another option would be to render a virtual keyboard (and hands) on top of the physical keyboard.
	
	Additional avenues for future work resides in probing the empirical user experience aspects of this work deeper by further experimentation. For example, our work raises questions about how to best design for perceived affordance, or how to best dynamically reconfigure a keyboard to assist users in complex workflows.
	
	\section{Conclusions}
	
	
	
	Physical keyboards are common peripherals for personal computers and are efficient standard text entry devices. While recent research has investigated how physical keyboards can be used in immersive HMD-based VR, so far, the physical layout of the keyboards has typically been directly transplanted into VR with the explicit goal of replicating typing experiences in a standard desktop environment.

	In this paper, we have explored how to fully leverage the immersiveness of VR to change the input and output characteristics of physical keyboard interaction within a VR environment. This allowed us to reconfigure the input and output mappings of both individual keys and the keyboard as a whole. We explored a set of input and output mappings for reconfiguring the virtual presentation of physical keyboards and designed, implemented and evaluated nine VR-relevant applications: emojis, languages and special characters, application shortcuts, virtual text processing macros, window manager, photo browser, a game (whack-a-mole), secure password entry and a virtual touch bar. We investigated the feasibility of the applications in a user study with 20 participants and found that the applications were usable in VR.
	
	From our results we see that we can integrate physical keyboards in VR experiences in many flexible ways. The biggest advantage of standard physical keyboards is that they are actually available as virtually every PC and laptop is already equipped with a physical keyboard. Instead of asking users to remove keyboards during VR use to make space for dedicated VR input devices, it might make more sense to flexibly integrate the keyboard into the VR experience, at least for scenarios such as office work \cite{grubert2018office}. Keyboards provide haptic feedback that can be used in many ways, for example for virtual keys, sliders, or to simulate reactive surfaces such as the case of whack-a-mole. They also provide accurate tactile guidance for the users' fingers.  For many input tasks, they are the fastest and most accurate input device available. 
	
	In conclusion, we have shown that physical keyboards can be used very flexibly as an input device for many different tasks in VR and could instantaneously reconfigure based on the context.  We believe that their unique advantages will make physical keyboards promising and flexible input devices for many VR experiences in the future.
	
	\balance{}

	\bibliographystyle{abbrv-doi}

	\bibliography{reconvigure}
\end{document}